\documentclass[smallextended]{svjour3}     
\usepackage{graphicx}%
\usepackage{amssymb}
\newcommand{\text}[1]{\hbox{\scriptsize\rm #1}}
\newcommand{\OmegaGW}{\Omega_{\text{gw}}}
\newcommand{\SGW}{S_{\text{gw}}}
\newcommand{\ShapeGW}{\mathfrak{S}_{\text{gw}}}
\newcommand{\prdref}[3]{Phys.\ Rev.\ D \textbf{#1}, {#2} (#3)}
\newcommand{\prlref}[3]{Phys.\ Rev.\ Lett.\ \textbf{#1}, {#2} (#3)}
\newcommand{\apjref}[3]{Astrophys.\ J.\ \textbf{#1}, {#2} (#3)}
\newcommand{\apjlref}[3]{Astrophys.\ J.\ Lett.\ \textbf{#1}, {#2} (#3)}
\newcommand{\cqgref}[3]{Class.\ Quant.\ Grav.\ \textbf{#1}, {#2} (#3)}
\newcommand{\roppref}[3]{Rep.\ Prog.\ Phys.\ \textbf{#1}, {#2} (#3)}
\newcommand{\aaref}[3]{Astron.\ Astrophys.\ \textbf{#1}, {#2} (#3)}
\newcommand{\lrrref}[3]{Liv.\ Rev.\ Rel.\ \textbf{#1}, {#2} (#3)}
\newcommand{\prslref}[3]{Proc.\ Roy.\ Soc.\ Lon.\ \textbf{#1}, {#2} (#3)}
\newcommand{\epjsref}[3]{Eur.\ Phys.\ J.\ ST \textbf{#1}, {#2} (#3)}
\newcommand{\physrepref}[3]{Phys.\ Rep.\ \textbf{#1}, {#2} (#3)}
\newcommand{\mnrasref}[3]{Mon.\ Not.\ R.\ Astron.\ Soc.\ \textbf{#1}, {#2} (#3)}
\newcommand{\appref}[3]{Astropart.\ Phys.\ \textbf{#1}, {#2} (#3)}

\journalname{General Relativity and Gravitation}

\begin{document}
\title{Current status of gravitational--wave observations}
\author{Stephen~Fairhurst, Gianluca~M~Guidi, Patrice~Hello,
John~T~Whelan, Graham~Woan}

\institute{Stephen Fairhurst \at
           Department of Physics and Astronomy\\
           Cardiff University, Cardiff, CF24 3AA, UK\\
           \email{stephen.fairhurst@astro.cf.ac.uk}
           \and
           Gianluca M Guidi \at
           Istituto di Fisica\\
           Via S Chiara 27, 61029, Urbino, IT\\
           Universita di Urbino, INFN Firenze\\
           \email{gianluca.guidi@uniurb.it}
           \and
           Patrice Hello\at
           LAL, Universite Paris-Sud, IN2P3/CNRS\\ 
           F-91898 Orsay, France\\
           \email{hello@in2p3.lal.fr}
           \and
           John T Whelan\at
           Center for Computational Relativity and Gravitation
           and School of Mathematical Sciences\\
           Rochester Institute of Technology,
           85 Lomb Memorial Drive, Rochester, NY 14623, USA\\
           \email{john.whelan@astro.rit.edu}
           \and
           Graham Woan\at
             Department of Physics \& Astronomy\\
             University of Glasgow\\
             Glasgow G12 8QQ,UK\\
             \email{graham@astro.gla.ac.uk}
}

\date{27 August 2009}

\maketitle
\begin{abstract}
The first generation of gravitational wave interferometric detectors
has taken data at, or close to, their design sensitivity.  This
data has been searched for a broad range of gravitational wave
signatures.  An overview of gravitational wave search methods and
results are presented.  Searches for gravitational waves from
unmodelled burst sources, compact binary coalescences, continuous
wave sources and stochastic backgrounds are discussed.
\end{abstract}
\section{Introduction}

The first generation of gravitational wave (GW) interferometric
detectors has reached an unprecedented sensitivity to GW signals.

Six interferometric detectors completed commissioning activities
and acquired scientific data over the last years.  Three of them
constitute the Laser Interferometric Gravitational-Wave Observatory
(LIGO), a joint Caltech-MIT project supported by the National
Science Foundation \cite{LIGO07} . They are situated in the USA, one
in Livingston, Louisiana (L1) and two, which share the same
facilities, in Hanford, Washington (H1, H2). L1 and H1 are
interferometers with arms of length 4 km, whereas H2 is a 2 km interferometer. Their
first Science Run began in September 2002.  Since then, four other
science runs took place; between November 2005 and September 2007
they operated at their design sensitivity in a continuous
data-taking mode in the 5th science run (S5).

\begin{figure}
\includegraphics[width=100mm]{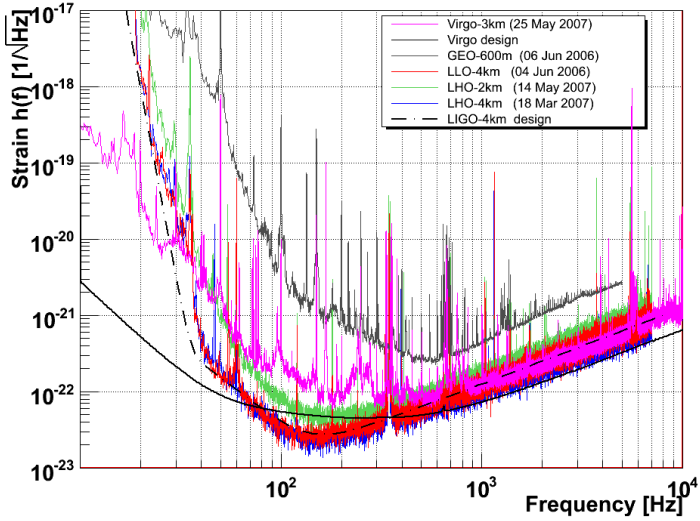}
\centering
\caption{\label{fig:Shh} Typical spectral density of calibrated noise for the
  LIGO interferometers and GEO during the S5 run and for Virgo interferometer
  during the VSR1
  run. Also displayed are the design sensitivities for the 4 km LIGO
  interferometers and for Virgo.}
\end{figure}

The Virgo detector is a joint project of the CNRS and INFN, operated
by the Virgo Collaboration at the European Gravitational Observatory
- a CNRS-INFN joint venture with the mandate to build the
interferometer \cite{VIRGO06}. Virgo is a 3 km interferometer near
Pisa (Italy). Its first Science Run (VSR1) took place in May
2007--September 2007, with a sensitivity comparable to LIGO instruments
above around 400 $\textrm{Hz}$ and a better sensitivity below 30 $\textrm{Hz}$. (see Fig.~\ref{fig:Shh}).

Two additional interferometric detectors are part of the network of GW
observatories: the Japanese TAMA project built a 300m interferometer near
Tokyo, Japan \cite{TAMA} and the German-British GEO project built a 600m
interferometer near Hannover, Germany \cite{GEO}; early in its operation GEO
joined with the LIGO Scientific Collaboration. The current sensitivity of TAMA and GEO is not
comparable to LIGO and Virgo, and the latest scientific
observational results are based on the analysis of S5/VSR1 data.

The LIGO S5 run collected a full year of triple detector coincidence
interferometer data, whereas Virgo ran in science mode for about 110
days during VSR1.  One of the most promising candidates for a GW
detection is thought to be the signal originated during the
coalescence of two neutron stars. The distance at which a system of
two neutron stars of 1.4 solar masses can be detected by a
gravitational wave observatory, with a signal to noise ratio of 8,
has thus become a standard measure of the detectors' sensitivity.
During the S5/VSR1 run, H1 reached a maximum sight distance of 35
Mpc, corresponding to a spherical observable volume of radius of 15
Mpc.

LIGO and Virgo signed an
agreement to jointly analyze data from the four LIGO-Virgo interferometers
collected after May 2007. The Data Analysis activity is run by four
LIGO-Virgo search physics groups with different scientific targets: Burst,
Compact Binary Coalescences, Continuous Waves and Stochastic
Background, each aiming at the detection and characterisation of
different sources of GWs. Even if this division is somehow arbitrary, it has
become the accepted standard in GW data analysis.

Currently, a new joint run, S6/VSR2, is on-going. The goal is to
improve the sensitivity of the detectors, Enhanced LIGO and Virgo+, by about a
factor 2 over the course of the run, resulting
in an observable volume of the universe of about an order of magnitude larger.

In the next sections of this article a general
introduction to the methods applied and observational results obtained for the
different targets will be presented.

\section{Bursts}

\subsection{Introduction}

Gravitational wave bursts are taken to be short signals, duration
$<$ 1s, with a large variety of possible waveforms, either wide or
rather narrow-band in the frequency domain. GW bursts historically
correspond to poorly modelled signals, due to the complex physics
involved in the description of the sources, such as core collapse
supernovae or mergers of two compact stars (black holes and/or
neutron stars).   Hence analysis methods seeking GW bursts must in
general stay robust with respect to the expected waveforms.
Moreover the benefit of running robust analysis methods is to stay
wide open to the unexpected, something which is fine when in a
discovery mode.  This constraint can be relaxed somewhat if some
additional information about the source is available from other
astrophysical observations.  Also, spectacular recent progress in
numerical relativity has allowed for accurate modelling of GW
sources.  For the most important burst sources we have now at least
an idea of what the emitted waveform looks like. 

\subsection{Sources of GW bursts}

Core collapse supernova --- the collapse of a massive star and
neutron star formation --- has long been envisaged as one of the
first GW sources.  Modern models now include 3D full relativistic
hydrodynamical simulations together with detailed micro-physics
\cite{Ott1,Ott2,Ott3,Gar}.  They predict the emission of GW signal
during the fast gravitational collapse and bounce with a duration $\sim1$
ms and peak amplitude $h \simeq 10 ^{-21}$ for sources located
at 10 kpc.  A longer (hundreds of ms) and in some cases
stronger signal is also expected after the bounce due to
hydrodynamical instabilities around the proto-neutron star.

Another source of GW bursts is the merger of two compact stars
(black hole and/or neutron star) which generally results in a black
hole, surrounded by a torus of matter if at least  one of the
progenitors is a neutron star. Progress in numerical relativity in
recent years allows a clear picture of the merger of two black holes
and a good prediction for the waveform \cite{pretorius,pretorius2}.
Good progress has been made too in modelling the merger of two
neutron stars \cite{rezzolla}, which is an even more complex problem
due to, for example, the role of the neutron star equation of state.

Core collapse supernova and merger of neutron stars are thought to
be the origin of, respectively, long and short Gamma Ray Bursts
(GRB).  In both cases, a black hole will be formed and surrounded by
a disk of matter.  The accretion of  matter onto the black hole
leads to the formation of relativistic jets and the emission of
gamma rays.  What is interesting in this situation is that a GW
signal and a GRB signal are emitted within a short delay. This may
considerably enhance the detection capability, as explained below,
and give deeper insight into the GRB mechanism.

Soft Gamma Repeaters (SGR) sporadically emit brief and intense
bursts of soft Gamma Rays and could also be a good source of GW
bursts. SGRs are produced by highly magnetized neutron stars which
undergo deformations (starquakes) that could excite the modes of the
star and then emit gravitational waves at about the same time as the
gamma emission. The expected waveform is typically a damped sinusoid
(``ringdown'') but the parameters depend on the equation of state of
the nuclear matter inside the star and are not accurately known.
Like in  the GRB case, the gamma ray emission permits a search for
the GW bursts in a narrow coincidence time window.

\subsection{Detection methods}
\label{ssec:burst_methods}

Based on the classification of expected sources given in the
previous section, there are basically two ways of searching for GW
bursts: a general one making very little or no assumption about the
signal direction, waveform, time of arrival, etc., and a second where
we take benefit of an external trigger (for example a GRB) for which
source location in the sky and timing are known.

The main steps of the analysis are trigger generation, efficiency
estimation and accidental background estimation, whose principles
are the same for both all-sky all-time unmodelled searches and
triggered searches.  Even with increasingly accurate predictions for
GW burst waveforms, it is still mandatory that burst algorithms
remain as robust as possible against the possible variety of
waveform. Indeed, generic signals such as Gaussian pulses or
Sine-Gaussian signals are used in parallel with astrophysical
waveforms (collapse, mergers) to test the efficiency and robustness
of trigger generators.  The up-to-date trigger generator algorithms
(or analysis  ``pipelines'') are time-frequency or equivalent
methods looking at local excess of energy (in time and frequency) in
the calibrated $h(t)$ time series
\cite{cwb1,cwb2,qpipeline,egc,xpipeline}.  These methods can be fully
coherent \cite{cwb1,cwb2,xpipeline} taking full benefit of the existing
network of detectors (LIGO and Virgo) or coincident with some
coherent follow-up when looking for possible event candidates. A
coherent method has the advantage that it can faithfully reconstruct
the burst signal and the position of the source in the sky. Of
course a coincident analysis can also reconstruct the sky location
of some burst event candidate. In both cases accurate sky position
can be reconstructed if the three LIGO and Virgo interferometers are
in operation and in science mode at the same time.  In the case of
the current LIGO-Virgo network, the typical position angular
accuracy is of the order a few degrees, depending on the GW
direction with respect to the detectors plane \cite{cavalier,marko}.

The efficiency estimation of the searches is performed by injecting
signals in the calibrated time-shifted output data of the detectors.
As already mentioned, these signals can be astrophysical signals
from core-collapse or merger simulations (see for example the search
for the merger and ring-down phases of binary black hole
coalescences with Virgo C7 commissioning run data \cite{C7Virgo}) or
generic signals. The so-called Sine-Gaussian signals (sinusoids with
Gaussian envelope) are particularly interesting since they can span
the detector bandwidth and give sensitivity estimates for the whole
range of accessible frequencies.  A number of such signals are then
added to the calibrated strain of each detector after time sliding.
For the all sky searches, they are generally distributed along the
entire data in order to average the detectors' noise
non-stationarities. For each injected signal the amplitude described
by the $h_{\mathrm{rss}}$, which corresponds to the total GW signal
energy:
\begin{equation}
h_{\mathrm{rss}} = \sqrt{ \int_{-\infty}^{+\infty} h(t)^2 \, dt} \,
.
\end{equation}
The $h_{\mathrm{rss}}$ amplitude has units of Hz$^{-1/2}$, meaning
we can compare it directly to the detector sensitivities, especially
when the injected waveform is well localized in frequency. The final
result for each signal is the efficiency curve, detection efficiency
vs. $h_{\mathrm{rss}}$. Usually the $h_{\mathrm{rss}}$ values at
50\% or 90\% efficiency are chosen for setting the upper limits.

Another important aspect of the burst analysis is the background
estimation and the rejection of loud outliers (glitches) using data
quality flags and event by event vetoes. The quality of the
detectors' data is studied at the individual detector level.  Short
duration noise transients can mimic a GW burst especially if their
rate is large enough to induce a number of coincident events between
the detectors. Data quality flags are used to indicate that one
interferometer is not working properly during some period, for some
known reason. Event-by-event vetoes are defined when an excess of
coincidence is found between the gravitational strain channel
triggers and auxiliary, environmental channel events.  The above
triggers are generally not obtained with the search pipelines but by
other algorithms which are required to be much faster since they run
on hundreds of auxiliary channels.  The procedure allows us to veto
only short time intervals (sub-second), thereby saving the
observation time of the detectors.  Once the data is cleaned by use
of data quality flags and vetoes it remains to estimate the
background, i.e., the number of accidental coincidences. As the
noise in each detector is independent, this is done by time shifting
the detector data with respect to the others and processing all
these time shifted data sets with the search algorithms. Relative
time shifts are chosen to be much larger than the light times of
flight between the detectors and much larger than the expected
signal durations.

There is a little difference for triggered searches. In this case,
the data are split into an on-source region where the GW burst is
expected and a background region where the noise is expected to be
statistically similar to the one in the on-source region. The
background region generally consists of a few hours from each side
of the on-source region. All the efficiency and background studies
are performed with the background data set.

All the cuts of the searches are then defined with these shifted
data sets and applied in a blind way to the original unshifted data
set. Events passing all the cuts (if any) are then detection
candidates and can be examined more closely. If no candidate is
found the result of the search can be turn into exclusion plots, for
instance rate of event vs. $h_{\mathrm{rss}}$ in the case of all-sky
analysis.  The frequentist upper limit for the rate of events with
confidence level $p$ is for example \cite{brady_etal}
 \begin{equation}
 R_p(h_{\mathrm{rss}}) = \frac{-\ln(1-p)}{T \epsilon(h_{\mathrm{rss}})}
 \end{equation}
where $T$ is the observation time and $\epsilon(h_{\mathrm{rss}})$
is the detection efficiency for the considered waveform with
amplitude $h_{\mathrm{rss}}$.

\subsection{Recent results}

The search for GW bursts has not yet yielded a detection and most
published results give upper limits on strain ($h_{\mathrm{rss}}$)
or rate.  When an astrophysical interpretation for a particular
source is possible then these limits can be converted into
meaningful astrophysical bounds.

The most recent all sky search for unmodelled GW bursts has been
completed using data from the first calendar year of the S5 LIGO
run.  The analysis has been split in two parts, a search for low
frequency signals in a 64-2000 Hz band \cite{S5y1LF} and a high
frequency search in a 1-6kHz band \cite{S5y1HF}. The observation
time $T \simeq 270$ days sets an upper limit about 3.6 events/year
for the rate of events at 90\% confidence level ($p=0.9$ in the
formula above) in the lower frequency band. In the higher frequency
band the network live-time is only $T \simeq 161$ days, resulting in
an upper limit for the rates of events about 5.4 events/year at 90\%
confidence level.  The strain limits depend on the details of the
injected signals for efficiency estimates. The typical $h_{rss}$
limit for Sine-Gaussian signals and most Gaussian pulses is below
$10^{-21}$ Hz$^{-1/2}$ in the low frequency band while it is up to a
few $10^{-20}$ Hz$^{-1/2}$ in the high frequency band.  Not
surprisingly the search sensitivity follows the detector
sensitivities which are better at low frequencies than at high
frequencies.

Another recent result concerns the search for a GW burst associated
with GRB 070201 \cite{GRB070201}, detected by gamma-ray satellites.
Those satellites found that the error box for the position of the
GRB is centered at about 1 degree from the center of M31 (Andromeda)
and overlaps the spiral arms of the galaxy.  Andromeda is the
closest spiral galaxy (at about 760 kpc) and an event possibly
occurring at such a short distance would be outstanding.  The GRB
was a short one and likely progenitors for short GRBs are binary
neutron star or neutron star-black hole coalescences or flares from
SGR. The analysis for coalescence waveforms is presented in the next
section.  An unmodelled search for a GW burst in association with
the GRB, yielded an upper limit for the radiated GW energy of about
$4.4\times 10^{-4}$ M$_\odot$c$^2$ for GW bursts lasting less then
100 ms for isotropic emission occurring at the LIGO peak sensitivity
near 150Hz.  This does not rule out the possibility of a SGR giant
flare in Andromeda galaxy.  Other searches for GW bursts associated
to GRB have previously been pubblished by the LIGO collaboration
\cite{GRBLIGO} and Virgo collaboration \cite{GRBVirgo}.

Two searches for GW bursts possibly associated with SGRs have been
recently published \cite{SGR,storm}. The first one targeted the SGR
1806-20 and the SGR 1900+14, including a giant flare episode of SGR
1806-20 which occurred  in December 2004 (during the LIGO
``Astrowatch'' period prior to the S4 run) and a storm episode of
SGR 1900+14 which occurred in March 2006 (during the LIGO S5 run).
Using a set of different waveforms the search was able to set upper
limits at 90\% confidence level on the isotropic GW emitted energies
in the range $3\times 10^{45}$ to $9\times 10^{52}$ ergs for a
source located at 10 kpc.  These upper limits depend on the detector
sensitivities and antenna patterns at the time of the Gamma
emission, on the loudest event in the on-source region and the
injected waveforms for efficiency estimation.  It is worth noting
that some theoretical models predict maximal GW emission energy as
high as $10^{49}$ ergs. This is well in the range of the SGR
analysis sensitivity.  The second paper is a re-analysis of the SGR
1900+14 storm of March 2006 with a different (``stacking'') method
\cite{storm}.  The gain in sensitivity is about one order of
magnitude with respect to the first analysis \cite{SGR}.

A more exotic analysis is the search for GW bursts emitted by cosmic
string cusps. The first reported results used the 2005 LIGO data (S4
run) \cite{cosmicstrings}. The search used matched filtering as the
predicted signals have simple and well parametrized waveforms.
Upper-limits have been set for the rate of events and for cosmic
string parameters such as string tension loop size or reconnection
probability. These limits are not competitive with the ones obtained
by other cosmological observations like indirect bounds from Big
Bang Nucleosynthesis, but analysis of the S5 data (much longer data
taking and with sensitivity twice as good) might surpass current
limits in a some portion of the cosmic string parameter space.

\section{Compact Binary Coalescence}

\subsection{The Binary Coalescence Waveform}

Coalescing binaries comprised of black holes and/or neutron stars
are ideal sources for gravitational wave detectors.  During the
latter stages of its evolution, a binary emits gravitational
radiation as the two component stars slowly spiral inwards before
finally merging to form a single object which settles down to
equilibrium.  Indeed, the emission of gravitational waves from
binary inspiral has been indirectly detected through observations of
binary pulsars \cite{weisberg:2004}, although the current
gravitational wave frequency is too low to be observed in
terrestrial gravitational wave detectors.  As the inspiral
progresses, however, the frequency and amplitude of the
gravitational waves increase.  During the final seconds or minutes
of inspiral and merger, the gravitational radiation emitted by
systems with a mass between one and several hundred solar masses
will lie in the sensitive band of ground based gravitational wave
detectors.

The precise form of the binary coalescence gravitational waveform
depends sensitively upon the parameters of the binary, most notably
the masses and spins of the binary components.  The eccentricity of
the orbit will affect the emitted waveform but, in most cases, it is
expected that the binary will have circularized before entering the
sensitive band of ground based detectors
\cite{Cokelaer:2009hj,Martel:1999tm}.  Historically, the binary
coalescence has been split into three parts: a slow inspiral, a
highly relativistic merger and ringdown to a final equilibrium
state.  Different techniques are used to calculate the waveform in
each of these regimes.  Depending upon the mass of the system,
different stages of the evolution will emit gravitational waves at
the sensitive frequency of the detector.

When the components of the binary are widely separated, the orbit
decays slowly due to energy emitted in gravitational radiation, and
the waveform sweeps slowly upwards in both frequency and amplitude.
During this inspiral phase, theoretical waveforms calculated within
the post-Newtonian framework \cite{Blanchet:2006av} are expected to
provide an accurate representation of the gravitational waveform.
The post-Newtonian waveforms derived to date are sufficient for
binaries comprised of neutron stars or low mass black holes (up to
about $10 M_{\odot}$), as the merger will occur above the most
sensitive frequency band of the detector.

It is expected that the end product of a binary system with total
mass above $2.0M_{\odot}$ will be a single, perturbed black hole
which rings down, by emission of gravitational radiation, to an
equilibrium conﬁguration.  Since stationary black holes are fully
characterized by their mass and angular momentum, the excitations of
higher multipoles, and in partiular the quadrupole, will be radiated
gravitationally \cite{2006PhRvD..73f4030B}.  The frequency and
damping time for each of the ringdown modes depends upon the mass
and angular momentum of the black hole  and can be calculated
analytically within the framework of black hole perturbation theory
\cite{Chandrasekhar:1975zz,Leaver:1985ax}.  Ringdown waveforms will
lie in the sensitive band of the detector for black holes with mass
greater than around $100 M_{\odot}$.

The dynamical merger of two black holes can only be modelled using
full general relativistic calculations.  Recent breakthroughs in
numerical relativity have, for the first time, enabled the
calculation of the gravitational waveform emitted during merger
\cite{pretorius2,pretorius}.  Currently, several groups are capable
of numerically evolving two black holes through their final orbits,
merger and ringdown
\cite{pretorius2,Husa:2007zz,Hannam:2009rd}.  The waveform for
binaries whose components have comparable mass, and are
non-spinning, is well characterized
\cite{Buonanno:2007pf,Damour:2007yf,Ajith:2007qp}.  There have also
been successes modelling the merger of neutron star binaries and
neutron star black hole binaries
\cite{rezzolla,Duez:2008rb,Read:2009,shibata:2007}.  Numerical
investigations of the full parameter space of compact binary
coalescence waveforms are ongoing.

\subsection{Search Methods}

As described above, the gravitational
waveform for coalescing binaries is well modelled analytically and
numerically.  Signal processing theory \cite{wainstein:1962}
advocates the use of matched filtering to extract known signals from
Gaussian noise.  Matched filtering provides the backbone of searches
for coalescing binaries \cite{FinnChernoff:1993,Allen:2005fk}.
However, two substantial challenges remain: searching over the large
parameter space of coalescing binary signals, and dealing with
non-stationarities in the data.

The full binary coalescence waveform depends upon as many as
seventeen parameters and it remains a challenge to search
efficiently over the full parameter space.  While some parameters,
such as the amplitude and coalescence phase of the waveform, can be
extracted using analytical techniques, others can only be searched
by repeatedly evaluating the matched filter at numerous points
across the parameter space.  This is facilitated by creating a bank
of template waveforms, subject to the condition that for any
candidate signal only a small fraction of the signal (typically 3\%)
is lost due to filtering with a mis-matched waveform
\cite{Owen:1995tm,Owen:1998dk,BBCCS:2006}. This method works well
for binaries with non-spinning components.  However, the parameter
space of spinning binaries is considerably larger.  Several methods
of attacking this problem have been proposed, including
phenomenological waveforms
\cite{BuonannoChenVallisneri:2003b,S3_BCVSpin}, restrictions to
binaries with a single spin \cite{Pan:2003qt} and Markov Chain Monte
Carlo \cite{vanderSluys:2007st,vanderSluys:2008qx} techniques.
However, the increase in both computational cost and the background
rate associated with covering the spin parameter space has, to date,
rendered this unfeasible.  Indeed, it was been shown that searching
for spinning binaries with non-spinning waveform templates provides
comparable sensitivity to the currently available spinning searches
--- the benefits of using the improved, spinning template model are
negated by the increase in false alarms, particularly in real data
\cite{VDBroeck:2008}.  Investigations of the use of spinning
templates in gravitational wave searches continue.

The data from gravitational wave interferometric detectors contains
a significant number of non-stationary transients caused by various
environmental and instrumental sources.  These reduce the
sensitivity of a matched filter search as loud noise transients will
produce a large signal to noise ratio, even if they do not match
well the gravitational wave binary coalescence signal.  Many
techniques have been developed to has mitigate the effect of these
noise transients in the data.  Among the most powerful are: data
quality and veto tests which flag times of poor data quality or use
auxiliary channels with known couplings to the gravitational wave
channel to remove times of poor data \cite{vetoGWDAW03}, as
described in Section \ref{ssec:burst_methods}; coincidence tests
which require that a signal be observed, with consistent parameters,
at widely separated sites \cite{Robinson:2008}; signal consistency
tests which compare the observed signal in the detector to the
predicted waveform \cite{Allen:2004,Rodriguez:2007};
the use of improved ranking statistics which better separate the
foreground and background by taking into account additional
information, over and above the signal to noise ratio of the
candidate \cite{LIGOS3S4all}.  By making use of these
additional tests, searches for binary coalescences are approaching
the theoretically predicted sensitivity in Gaussian noise.

\subsection{Search Results and Future Prospects}

Gravitational wave data from the GEO, LIGO, TAMA and Virgo
interferometric detectors have been analyzed for coalescing binary
signals.  To date, no gravitational wave signal has been observed.
Consequently an ever improving set of upper limits has been placed
on the rate of binary coalescence as a function of the mass of the
binary.  Upper limits have been derived for systems ranging from
neutron star binaries through to intermediate mass black hole
binaries.  Here, we recap the latest results and compare them
with astrophysical predictions.

Astrophysical estimates of binary neutron star coalescence rates can
be derived from observations of binary pulsars in the galaxy.  These
rates are extrapolated to the local universe under the assumption
that the rate of binary coalescence follows the star formation rate
in spiral galaxies, which is obtained from measurements of the blue
light luminosity of galaxies \cite{LIGOS3S4Galaxies}.  Thus, results
are quoted per $L_{10}$ per year, where $1 L_{10} = 10^{10}$ times
the solar blue luminosity and, for reference, the Milky Way is
approximately $1.7 L_{10}$.  The predicted rate for binary neutron
star coalescence is $5 \times 10^{-5} L_{10}^{-1} \mathrm{yr}^{-1}$,
although the rate could plausibly be as much as an order of
magnitude larger \cite{Kalogera:2004tn}.  To date,
there are no direct observations of black hole-neutron star or black
hole-black hole binaries.  Thus, rate estimates are based upon
population synthesis, and yield realistic rates of $2 \times 10^{-6}
\textrm{ yr}^{-1} L_{10}^{-1}$ for neutron star-black hole binaries
\cite{Oshaughnessy:2008} and $4 \times 10^{-7} \textrm{ yr}^{-1}
L_{10}^{-1}$ for binary black holes \cite{OShaughnessy:2005}.  In
both cases, the rate could feasibly be as high as $6 \times 10^{-5}
\textrm{ yr}^{-1} L_{10}^{-1}$.

The first eighteen months of the LIGO S5 data have been analyzed for
gravitation wave signals from coalescing binaries with a total mass
less than $35 M_{\odot}$
\cite{Abbott:2009tt,Collaboration:2009lm1218}.  Upper limits
obtained by combining the result of this analysis with results from
S3 and S4 provide the most stringent bounds on the coalescence rate
from gravitational wave observations.  The binary neutron star
coalescence rate is restricted, at 90\% confidence, to be less than
$1.4 \times 10^{-2} L_{10}^{-1} \mathrm{yr}^{-1}$
\cite{Collaboration:2009lm1218}.  This is a factor of thirty above
optimistic rates, and several hundred above the best estimate of the
rate.  It is, however, interesting to note that the upper limit has
improved by four orders of magnitude from the one obtained with
LIGO's first science run \cite{LIGOS1iul}.  For binary black holes,
with component masses $5 \pm 1 M_{\odot}$, the 90\% rate limit is $9
\times 10^{-4} L_{10}^{-1} \mathrm{yr}^{-1}$, and for black
hole-neutron star binaries, the limit is $4 \times 10^{-3}
L_{10}^{-1} \mathrm{yr}^{-1}$ \cite{Collaboration:2009lm1218}.
These limits are again between one and two orders of magnitude from
the upper end of astrophysical predictions, and three orders of
magnitude from best estimates.

The most recent upper limits for binaries with a total mass greater
than $35 M_{\odot}$ were obtained using data from the fourth LIGO
science run.  For binaries of total mass between $30$ and $80
M_{\odot}$ a search with a phenomenological template family
\cite{BuonannoChenVallisneri:2003a} of binary black hole waveforms
gave an upper limit of $\sim 1 L_{10}^{-1} \mathrm{yr}^{-1}$
\cite{LIGOS3S4all}.  For higher masses, a search for the ringdown
portion of the signal yielded a rate limit for binary coalescences
in the mass range $100$ to $400 M_{\odot}$ of $1.6 \times 10^{-3}
L_{10}^{-1} \mathrm{yr}^{-1}$.  A search of the LIGO S5 data for
black hole binaries with a total mass up to $100 M_{\odot}$ is being
pursued \cite{Robinson:2009}.  This search will, for the first time,
make use of full inspiral-merger-ringdown coalescence waveforms
obtained by enhancing the post-Newtonian inspiral waveforms with
merger and ringdowns simulated numerically \cite{Buonanno:2007pf}.

The coalescence of two neutron stars or a neutron star and a black
hole is one of the preferred progenitor scenarios for short-duration
GRBs \cite{nakar07}.  By making use of the known time and sky
location of observed GRBs, it is possible to perform a more
sensitive search of the gravitational wave data.  This has been done
for GRB 070201, which was a short GRB localized in a region of the
sky which overlapped the Andromeda galaxy \cite{GRB070201}.  The
search yielded no evidence of gravitational waves, and allowed for
the exclusion of a binary coalescence progenitor in M31 with 99\%
confidence.

Recently data taking with the enhanced LIGO and Virgo+ detectors
began.  It is hoped that these detectors will achieve a factor of
two sensitivity improvement over the initial detectors, which
translates to almost an order of magnitude increase in the volume of
the universe that can be probed for binary coalescences.  In the
following years, advanced gravitational wave detectors will bring an
order of magnitude increase in sensitivity over the initial
configurations.  At this stage, astrophysical estimates predict the
observation of gravitational waves from tens of binary coalescences
per year.

\section{Continuous Waves}
\subsection{Introduction}
Although neutron stars in coalescing binary systems represent a
relatively well-understood population of putative gravitational
wave sources, with well-defined gravitational luminosities and
population statistics, the same neutron stars (and even isolated
neutron stars) can in principle radiate a detectable amount of
gravitational radiation well before coalescence.  Radio and X-ray
pulsar populations give us only a hint of the vast number
($\sim10^9-10^{10}$) of neutron stars that exists in the Galaxy. We
currently see perhaps only one in a million neutron stars as a
pulsar, but any neutron star, pulsar or not, can generate
continuous quasi-sinusoidal gravitational waves through rotation.

\subsection{Sources}
Any non-axisymmetric spinning neutron star will generate
gravitational radiation. Although the centrifugal deformation  can
be expected to make it significantly oblate ($\sim 10^{-4}$), this
axisymmetric deformation will not itself generate gravitational
radiation. Instead one requires the shape of the neutron star to be
supported against relaxation to a fluid equipotential surface by a
force, possibly an elastic stress force from the crust of the star, a magnetic
force distorting the crustal shape or possibly a distortion caused
by accretion or gravitational radiation-driven instabilities. A neutron star with
its spin axis oriented towards us, at a distance $d$, with such a
mass quadrupole moment $Q$ around its axis of spin will generate a
circularly polarised gravitational signal, at a frequency equal to
twice the rotation rate $\nu$ of the star, with amplitude
\begin{equation}
 h_0 = \frac{16G\pi^2}{c^4}\frac{\nu^2}{d}Q.
\end{equation}
If the spin axis is inclined to the line-of-sight by an angle
$\iota$, the radiation becomes elliptically polarised and the
amplitude is reduced.  It is often convenient to express $Q$ as the
product of an axial moment of inertia, $I_{zz}$, and an effective
equatorial ellipticity,
\begin{equation}
\epsilon = \frac{I_{xx}-I_{yy}}{I_{zz}},
\end{equation}
where $I_{xx}$ and $I_{yy}$ are the other two principal moments of
inertia.  There are clearly two very important questions here: i)
what is the size of deformation that neutron stars \emph{could}
have, given their equation of state, crystalline structure and
physical environment, and ii) what deformations do they
\emph{actually} have.  The answer to the first question depends
critically on both the equation of state of the neutron star (and
whether it is indeed a neutron star or a quark star) and its
crystalline structure \cite{UCB00,Owen05}.  Recent work by Horowitz and
Kadau \cite{HK09} has indicated that the structure may indeed by
highly crystalline, with point defects rapidly squeezed out to give
breaking stains of as much as 0.1.  This would allow self-supported
equatorial ellipticities of perhaps $10^{-5}$ to $10^{-6}$.

This second question is harder to address. Unlike in binary neutron
star systems such as PSR B1913+16, PSR J0737$-$3039, PSR B1534+12
and PSR J1756$-$2251, where the orbital evolution presents
convincing evidence of the very early stages of a coalescence, we
have no direct evidence of spin-gravitars (that is, neutron stars
whose observed spin-down is well-modelled by gravitational
braking). Certainly in the case of equatorial deformation supported
by crustal strength we cannot dismiss the notion that some neutron
stars have perfectly annealed equipotential surfaces, with
negligible axial quadrupole moment and therefore essentially no
gravitational luminosity.  For example, the extremely low period
derivatives of millisecond pulsars hint that these neutron stars at
least show very little equatorial asymmetry. At some level all
neutron stars will show deformation due to internal magnetic 
pressures, though this only becomes relevant for the strongest of 
magnetars \cite{Col08}. However, the story is
less clear for young and/or accreting neutron stars.  For both
these classes there are plausible mechanisms to supply both the
energy and the deformation necessary for significant gravitational
luminosity (see for example \cite{FS78,Bil98,Owen06}).

For an isolated neutron star we may postulate that the
gravitational luminosity should be less than the rate of loss of
rotational kinetic energy for a rigid body. In turn this defines an
upper limit on the strain amplitude we could expect from rigid body
gravitar with a spin-down rate $\dot\nu$ as
\begin{equation}
h_0 \leq \left(\frac{5GI_{zz}}{2c^2d^2}\frac{|\dot\nu|}{\nu}\right)^{1/2}.
\end{equation}
This ``spin-down upper limit'' falls below the 1-year strain
sensitivity of both the initial LIGO and Virgo detectors for all
but a small number of known radio pulsars, making these pulsars
unlikely first-detection candidates.  However, only a tiny fraction
of the neutron star population is seen as electromagnetic pulsars,
leaving the possibility of a gravitationally luminous, but
electromagnetically dim, population.

\subsection{Search methods}
There is an obvious sense in which it is easier to search for a
continuous quasi-sinusoidal signal than a transient inspiral or
burst signal.  A long-lived signal can be re-observed (or even
retrospectively observed using archived data) and confirmed as
astrophysical. In addition, there are radio and X-ray pulsars that
accurately trace the rotational evolution of around 100 pulsars
whose gravitational signal would fall in terrestrial observing
bands.  However, the benefits of a continuous wave search stop
there.  An inevitable consequence of searching for a long-duration
deterministic signal containing up to $\sim 10^{10}$ cycles is an
exquisite sensitivity to its parameter values.  Most obviously, a
change in frequency corresponding to just one more or one fewer
cycles during the observation would represent an entirely different
search, with a template for the expected signal that, as a matched
filter,  was insensitive to the original signal.  It also becomes
apparent that if the neutron star is spinning down, the number of
spin-down templates necessary to cover the possible alternatives
scales as the square of the observing time. Additionally, the
doppler modulation of the received signal is sensitive to the
position of the source on the sky.  For year-long observations this
angular sensitivity is approximately the gravitational diffraction
limit of an aperture the diameter of the Earth's orbit about the
Sun ($\sim 1$\,arcsec at typical frequencies). If we do not have
the benefit of a radio trace of the neutron star's rotational
evolution and sky location we are forced to perform a search over
this parameter space, and it rapidly becomes apparent that the
parameter space is huge. Continuous-wave searches are by far the
most computationally expensive searches that the gravitational wave
community undertakes, and in its most general form the problem is
(and always will be) fully limited by available computing power.

No matter what its form, any coherent search method will improve
its strain signal-to-noise ratio as
\begin{equation}
{\rm snr}\propto (S_h/T)^{-1/2},
\end{equation}
where $S_h$ is the detector's (strain) power spectral density at
the frequency of the signal and $T$ is the observing time. However
the overall sensitivity of a search is not solely dependent on
signal-to-noise ratio.  The more trials that are undertaken (i.e.,
templates that are searched) the greater the probability of random
noise popping up to unluckily appear like a signal.  The apparent
signal-to-noise ratio indicative of a true signal is therefore
several tens for searches that pick the strongest candidate from
over a wide parameter space. One consequence of this is that any
convincing signal must have a relatively high signal-to-noise ratio
after only a relatively short coherent integration.  This allows
one to combine these short integrations incoherently (as powers,
ignoring phase) without too great an impact in overall sensitivity
and develop semi-coherent search methods which are computationally
much cheaper \cite{CGB05}.  The overall sensitivity does however
only improve as the quarter power of the number of incoherently
combined contributions.

Ground-based CW search efforts have concentrated on variants of the
above, from fully coherent long-timescale searches for
gravitational wave signals phase-locked to radio pulsars
\cite{A04,A05b,A07b,A08b} to searches concentrating on non-pulsing
targets \cite{A07a,A07c} and massively computational all-sky
searches using a variety of semi-coherent techniques
\cite{A05a,A08a,A09a,A09b,A09c}.

\subsection{Search results}
Upper limits on the strength of continuous gravitational waves from
both known and unknown galactic neutron stars have been made
regularly since the first LIGO/GEO science run in 2002. As
sensitivities and run lengths have improved, the limits have
steadily dropped. The recent 23-month LIGO S5 run had sufficient
sensitivity to show that the Crab pulsar is not a gravitar (i.e.,
is not spinning down solely due to the emission of gravitational
radiation).  In itself this is no surprise -- the overall energy
budget of the Crab nebula and pulsar has to account for the nebula
luminosity and expansion. However, the early S5 result (covering
just the first 9 months of data) was sufficiently sensitive to show
that less than about 3 percent of the spin-down luminosity of the
Crab pulsar is due to gravitational emission \cite{A08b}. The full
S5 result, with a fully coherent search for gravitational emission
from 116 known pulsars, including the Crab pulsar, is expected
soon. Additional work is going on to use these and other
newly-developed targeted algorithms to search Virgo VSR1 data for
emission from the Vela pulsar at $\sim 22.5$\,Hz \cite{vela}.

The S5 run has also resulted in the most sensitive ``all
sky'' (i.e., survey) searches to date. Two early-S5 papers have
already been published on this \cite{A09b,A09c}.  The first
comprised a semi-coherent search, incoherently adding 30-minute
demodulated power spectra (the ``power flux method'').  The second
was an early result from Einstein@Home, a distributed screen saver
application that currently attracts about 200\,000 users worldwide
and returns $\sim100$\,Tflops to search project.  This method
looked for coincident detections between multiple  30-hour coherent
searches. Both these all-sky searches  returned strain upper limits
of around $10^{-24}$ for a wide spectral band, and these are levels
with real astrophysical significance. A simple argument, originally
by Blanford but developed by Knipsel and Allen \cite{KA08} indicates
that for our Galaxy population and distribution of neutron stars,
the loudest expected CW source would have a strain at Earth at
about this level (under certain assumptions).  Such a source would
need to be within a few hundred parsecs of Earth.  In addition to
targeted and all-sky searches, more specialised ``directed'' search
methods are also being used to tackle likely sky locations
including globular clusters, the low-mass X-ray binary Sco-X1, the
galactic centre and supernova remnants. One such search, for
gravitational emission from the X-ray point source at the centre of
Cas~A is nearing completion and has reported an expected
sensitivity also in the range $\sim10^{-24}$ \cite{wette08}.
The Ligo and Virgo Collaborations have now developed a broad suite
of algorithms and methods to tack a wide range of potential sources
of continuous gravitational radiation, including all-sky searches
for binary sources, and the full power of these will be applied to
data form the current S6/VSR2 runs.

\subsection{Future prospects}
As with other searches that involve population statistics, the
crude extrapolation holds that a factor $\eta$ improvement in
sensitivity will increase detection numbers by a factor $\sim
\eta^3$.  Clearly the physical extent of the Galaxy places an upper
limit on this, but that only becomes relevant for current all-sky
searches when broadband sensitivities are a factor $\sim100$ times
their current values. Perhaps more important is a consideration of
the types of neutron star that may be detectable in the future
using instruments with an improved low frequency response. Current
detectors show good sensitivity only to relatively rapidly spinning
pulsars, most of which are recycled millisecond pulsars with low
observed spin-down rates and, probably, low gravitational
luminosity. Young, glitchy pulsars are more common at gravitational
frequencies below $\sim100$\,Hz, with some of the most interesting,
rapidly braked, sources closer to 10\,Hz, so the low-frequency wall
is a particular challenge for future continuous wave gravitational
observations.

\section{Stochastic Background}

A stochastic gravitational-wave background (SGWB) refers to a
long-lived random GW signal.  This is generally produced by a
superposition of many unresolved sources, and can be characterized as
cosmological or astrophysical according to the epoch in which the GWs
are generated.  Cosmological backgrounds can be assumed to be
approximately isotropic, unpolarized and stationary, while
astrophysical backgrounds may have additional structure depending on
the nature of their sources.

\subsection{Sources}

A convenient measure of the strength of a SGWB is the energy density
in the GWs, per logarithmic frequency interval, in units of the
critical energy density needed to close the universe:%
\footnote{Note that $\rho_{\text{crit}}=(3H_0^2c^2)/(8\pi G)$
  depends on the value of the
  Hubble constant; it has become conventional to use the fiducial
  value $H_0=72\,\textrm{km}/\textrm{s}/\textrm{Mpc}$ when defining
  $\OmegaGW(f)$.}
\begin{equation}
  \OmegaGW(f) = \frac{1}{\rho_{\text{crit}}}
  \frac{d\rho_{\text{gw}}}{d\ln f}
\end{equation}
Cosmological models which produce a SGWB include amplification of
quantum vacuum fluctuations during inflation
\cite{Grishchuk:1975,Grishchuk:1997,Starobinsky:1979}, phase
transitions \cite{Kosowsky:1992,Apreda:2002}, pre-big-bang models
\cite{Gasperini:1993,Gasperini:2003,Buonanno:1997}, and cosmic
(super-)string models
\cite{Caldwell:1992,Damour:2000,Damour:2005,Siemens:2008}.  Standard
inflationary models generate a backround of constant $\OmegaGW(f)$
over many decades of frequencies, but the amplitude of such a
background is already bounded by cosmic microwave background
observations to be $\OmegaGW(f) < 10^{-14}$ \cite{Smith}. Astrophysical
GW backgrounds can be generated by unresolved superpositions of
sources such as cosmic string cusps \cite{Siemens:2008},
supernovae \cite{Coward:2002}, and neutron-star
instabilities \cite{Regimbau:2001,Regimbau:2006}.

The most stringent indirect limit on a SGWB in the frequency range of
ground-based detectors comes from a constraint on the total energy
density present at the time of nucleosyntheis.  This big-bang
nucleosynthesis (BBN) bound limits the total energy density in
gravitational waves to be
\begin{equation}
  \label{e:BBNbound}
  \int \frac{df}{f}\OmegaGW(f) \lesssim 1.1\times 10^{-5} (N_\nu -3)
\end{equation}
where the effective number $N_\nu$ of neutrino species at BBN is 
constrained to by $(N_\nu -3)<1.44$.\cite{Cyburt:2005}
Note that this limit only applies to cosmological SGWBs, i.e.,
gravitational waves generated before the era of nucleosynthesis.

\subsection{Search Methods}

Since the amplitude of a SGWB will be much smaller than that of
instrumental noise in a typical ground-based detector, one needs to
exploit the expectation that while instrumental noise will be
(predominantly) uncorrelated between independent detectors, the
gravitational wave signals in a pair of detectors should have an
average correlation
\begin{equation}
  \langle \tilde{h}_1(f)^* \tilde{h}_2(f') \rangle
  = \frac{1}{2}\delta(f-f') \gamma_{12}(f) \SGW(f)
\end{equation}
where $\gamma_{12}(f)$ encodes the observing geometry (location and
orientation of detectors 1 and 2, and in the case of an anisotropic
background, the spatial distribution of the background) and
$\SGW(f)$ is a one-sided power spectral density for the SGWB
which is given for an isotropic background by
\begin{equation}
  \label{SGW}
  \SGW(f)=[(3H_0^2)/(10\pi^2)]f^{-3}\OmegaGW(f)
  \ .
\end{equation}

The standard search method \cite{AR99} for an isotropic background
cross-correlates the data from pairs of detectors using an optimal
filter
\begin{equation}
  \tilde{Q}(f) \propto
  \frac{\gamma_{12}(f)\ShapeGW(f)}{S_1(f)S_2(f)}
\end{equation}
where $S_{1,2}(f)$ are the noise power spectra for the two detectors
and $\ShapeGW(f)$ is the expected shape of
the SGWB spectrum.  The resulting search is sensitive to a background
$\SGW(f)=S_R \ShapeGW(f)$ of strength
\begin{equation}
  S^{\text{detectable}}_R \sim
  \left(
    2T
    \int_{0}^{\infty}
    df\,
    \frac{\left[\gamma_{12}(f)\,\ShapeGW(f)\right]^2}{S_1(f)\,S_2(f)}
  \right)^{-1/2}
  \ .
\end{equation}
Note that the sensitivity of a cross-correlation search improves like
the square root of the observing time $T$.  Also, stochastic
background measurements tend to be dominated by the low end of the
available frequency range, because $\gamma_{12}(f)$ oscillates with
increasing $f$ within an envelope whose leading term is $\propto f^{-1}$
and because Eq.~\ref{SGW} means that a constant-$\OmegaGW(f)$
background has $\ShapeGW(f)\propto f^{-3}$.

A cross-correlation search can also be used to search for an
astrophysical background with a specified spatial distribution, e.g.,
a SGWB coming from one point on the sky \cite{Ballmer06}. More
sophisticated techniques can be used to recover the spatial
distribution of a measured background \cite{Mitra08}.

\subsection{Search Results}

The most stringent direct limit on $\OmegaGW(f)$ was set using data
from the S5 run of LIGO Livingston and LIGO Hanford \cite{S5iso},
which set the 95\% confidence level upper limit of
$\OmegaGW(f)<6.9\times 10^{-6}$ assuming $\OmegaGW(f)$ to be constant
over the interval $41.5\,\textrm{Hz}<f<169.25\,\textrm{Hz}$.  A SGWB
of the excluded strength, confined to those frequencies, would
contribute $9.7\times 10^{-6}$ to the total value of $\Omega$.  This
limit is therefore more stringent than the BBN bound
(Eq.~\ref{e:BBNbound}) and we have entered the era where ground-based
GW detectors are placing new limits on gravitational wave backgrounds
of cosmological origin.

The previous limit from S4 data \cite{S4iso}, while less stringent by
about an order of magnitude, already placed new restrictions on the
parameters of some cosmic string models which generate GWs both before
and after the era of nucleosynthesis.
Additional searches of S4 LIGO data set limits on the strength of
possible point-like backgrounds \cite{S4map} and (by correlating LIGO
Livingston data with data from the ALLEGRO bar detector) set a
higher-frequency limit of
$\OmegaGW(915\,\textrm{Hz})<1.02$ \cite{S4AL}.
Correlation measurements using LIGO and Virgo data are expected
to improve the high-frequency measurement \cite{Cella07}. Further
searches for anisotropic backgrounds are also being conducted.

\section{Discussion}

The current search for GW covers multiple types of signals
originating from different possible astrophysical events like core
collapse of massive stars and neutron star formation, coalescing
binary systems of neutron stars and black holes, non-axysimmetric
spinning neutron stars and signals produced by a large collection of
incoherent sources. The data acquired by the most sensible GW
observatories, which are at present the LIGO and Virgo
interferometers, are analysed applying different methods and
strategies targeted to the identification and characterisation of
the signals emitted by these possible sources. Moreover, methods
able to catch signals coming from unknown sources are also currently
used.

The analysis of the latest scientific data, acquired during the
first part of the S5/VSR1 run, did not provide any evidence of a
possible detection. Upper limits on the rate of events and/or the
strain amplitude $h$ are then derived and compared to the
astrophysical predictions. These limits are already valuable
scientific results which reinforce and widen our knowledge of the
astrophysical events involved.  For example they are now approaching
the plausible astrophysical values in the case of GW originating
from binary coalescing systems and have already set a bound to the
percentage of the spin-down luminosity of the Crab pulsar on the
energy emitted in gravitational radiation.
The energy density in a stochastic GW background around
100~$\textrm{Hz}$ has been constrained to a limit which is more
stringent than the the big-bang nucleosynthesis bound, the strongest
indirect limit at those frequencies.
From the analysis of the data in
coincidence with the GRB 070201 it has been possible to exclude the
hypothesis of a binary merger in M31 as the progenitor of this
event.

Further results from the analysis of the full S5/VSR1 run data are
expected soon. Meanwhile, the analysis of the data from the S6/VSR2
run will probably, if no GW detections are found, improve the
current limits, owing to the possible improvement of sensitivity of
LIGO and Virgo detectors.  The sensitivity of the present and future
GW detectors gives the possibility of studying astrophysical events
jointly with other observatories, i.e.  electromagnetic and neutrino
observatories, likely bringing additional information on the physics
of the sources and on their characteristics. The future searches
will then open the possibility of performing a mature GW astronomy.

\section*{Acknowledgements}

The authors would like to thank their colleagues in the LIGO and
Virgo Scientific Collaborations, especially Ray Frey, Ben Owen and
Joe Romano for comments and suggestions.  SF would like to
acknowledge the support of the Royal Society.  JTW is supported by
NSF grant PHY-0855494 and by the College of Science of Rochester
Institute of Technology.

\end{document}